\documentclass[a4paper]{jpconf}
\usepackage{graphicx}
\usepackage{citesort}
\usepackage{lineno}
\usepackage{amsmath}
\begin{document}
\title{Azimuthal correlations of the longitudinal structure of the mid-rapidity charged-particle multiplicity in Pb-Pb collisions at \ensuremath{\sqrt{s_{\mathrm{NN}}}} = 2.76 TeV with ALICE}
\author{Saehanseul Oh (for the ALICE Collaboration)}
\address{Wright Laboratory, Yale University \\ 266 Whitney Avenue, New Haven, CT 06511, USA}
\ead{saehanseul.oh@yale.edu}

\begin{abstract}
Studies of longitudinal correlations of the charged-particle multiplicity in heavy-ion collisions have provided insights into the asymmetry and fluctuations of the initial-state collision geometry. 
In addition to the expansion of the medium in the transverse direction, commonly quantified using Fourier coefficients ($v_{n}$), the initial  geometry and resulting longitudinal expansion as a function of azimuthal angle enable us to better understand the full 3-dimensional picture of heavy-ion collisions. 
In these proceedings, azimuthal correlations of the longitudinal structure of charged-particle multiplicity are reported for Pb-Pb collisions at a nucleon-nucleon center-of-mass energy of 2.76 TeV. 
The azimuthal angle distribution is divided into regions of in-plane and out-of-plane with respect to the second-order event plane, and the coefficients of Legendre polynomials are estimated from a decomposition of the longitudinal structure of the charged-particle multiplicity at midrapidity ($|\eta| < 0.8$) on an event-by-event basis in each azimuthal region for different centralities. 
Correlations between the coefficients of various orders in different azimuthal regions are studied and exhibit collective features of longitudinal structure in the azimuthal direction. 
The results are compared with HIJING and AMPT simulations. 
\end{abstract}

\section{Introduction}
Collective flow is one of the key features in relativistic heavy ion physics. 
The initial state anisotropy and its fluctuations are translated into the anisotropy in the final state particle distribution.
This has been investigated extensively in the transverse direction via the Fourier decomposition of particle distributions and its coefficients, $v_{n}$~\cite{Aamodt:2010pa}. 
Similar studies in the longitudinal direction were proposed recently~\cite{Bzdak:2013aa,Bzdak:2016ec}, where results from the ATLAS Collaboration indicate different levels of contributions for short-range and long-range correlations in particle distributions in the longitudinal direction for various collision systems~\cite{ATLAS:2017aa}.
While these approaches are limited either to the transverse or longitudinal direction, there are studies aimed to extend the analysis to both azimuthal and longitudinal directions, such as measurements of pseudorapidity ($\eta$) dependent $v_{n}$~\cite{CMS:2017aa,ATLAS:2012330} and the decorrelation of the transverse flow in the longitudinal direction~\cite{ATLAS:2018fd}.
Using the decomposition in the transverse direction as a basis, they reveal hints of an interplay between transverse and longitudinal dynamics.  
Such measurements are essential to understand the full 3-dimensional expansion of the medium created in heavy ion collisions.  

The ALICE Collaboration has recently devised new observables in order to further investigate 3-dimensional features using particle correlations. 
The decomposition in the longitudinal direction is used as a basis, and correlations among longitudinal structures in different azimuthal regions are studied. 
In the present proceedings, these new observables will be introduced, and the corresponding results in Pb-Pb collisions at a nucleon-nucleon center-of-mass energy of 2.76 TeV at the LHC will be shown.

\section{Definition of observables}
\label{Sec:Def}
In a similar manner to that of the decomposition of particle distributions in azimuthal angle ($\varphi$) using Fourier harmonics, particle distributions in the longitudinal direction can be  decomposed using Legendre polynomials~\cite{ATLAS:2017aa}.
Adding azimuthal information in the definition, the single particle density ratio within a limited azimuthal ($\varphi$) range, $R_{\varphi_{i}}(\eta)$, is defined by 
\begin{eqnarray}
R_{\varphi_{i}}(\eta) \equiv \frac{1}{\int_{\varphi_{i}} \text{d}\varphi} \int_{\varphi_{i}} \frac{N(\varphi, \eta)}{\langle N(\varphi, \eta) \rangle} \text{d}\varphi\; \text{,}
\label{Eq:NRatio}
\end{eqnarray}
where the integral range denoted by $\varphi_{i}$ is equal to the range of the corresponding $\varphi$ bin, $N(\varphi,\eta)$ is the multiplicity distribution as a function of $(\varphi, \eta)$ in a single event, and $\langle\rangle$ corresponds to the average within the given centrality class.  
The advantage of using $R_{\varphi_{i}}(\eta)$ instead of $\text{d}N/\text{d}\eta$ within a limited $\varphi$ range includes larger sensitivity to the event-by-event variations, a reduced influence from finite tracking efficiencies and related systematic uncertainties.
For the range of $\varphi_{i}$ in Eq.~\ref{Eq:NRatio}, the azimuthal angle is divided into four regions with respect to the second-order event plane ($\Psi_{2}$) in each event, i.e. two in-plane regions ($\varphi_{\text{in}}$) and two out-of-plane regions ($\varphi_{\text{out}}$).  
Two $\varphi_{\text{in}}$ bins cover the $\varphi$ ranges $\Psi_{2} - \frac{\pi}{4} < \varphi < \Psi_{2} + \frac{\pi}{4}$ and $\Psi_{2} +\frac{3\pi}{4} < \varphi < \Psi_{2} + \frac{5\pi}{4}$, and $\varphi_{\text{out}}$ bins cover $\Psi_{2} + \frac{\pi}{4} < \varphi < \Psi_{2} + \frac{3\pi}{4}$ and $\Psi_{2} +\frac{5\pi}{4} < \varphi < \Psi_{2} + \frac{7\pi}{4}$. 

Within the considered $\eta$ range $[-Y, Y]$, $R_{\varphi_{i}}(\eta)$ is then decomposed as 
\begin{eqnarray}
R_{\varphi_{i}} (\eta) \propto 1 + \sum\limits_{n=1} a_{n}T_{n}(\eta) \; 
\label{eq:Rdecomp}
\end{eqnarray}
with
\begin{eqnarray}
T_{n}(\eta) = \sqrt{n + \frac{1}{2}} P_{n}(\eta/Y) \; \text{,}
\label{Eq:Decomp}
\end{eqnarray}
where $P_{n}$ is the $n$-th order Legendre polynomial and $\sqrt{n + \frac{1}{2}}$ is a normalization factor, such that  
\begin{eqnarray}
\int_{-Y}^{Y} T_{n}(\eta)T_{n}(\eta)\,\text{d}\eta =1 \; \text{.}
\end{eqnarray}
The first few terms of Legendre polynomials are $P_{0}(x) = 1$, $P_{1}(x) = x$, $P_{2}(x) = (3x^2 - 1)/2$.
The coefficients $a_{n}(\varphi_{i})$ of the decomposition represent the longitudinal structure within a given $\varphi_{i}$ range and centrality class.

Once the $a_{n}(\varphi_{i})$ coefficients are extracted in each event, correlations among the coefficients can be used to describe the correlations among the longitudinal structures in different azimuthal regions. 
These correlations are studied through the conditional $a_{n}$, denoted by $a_{n}(\varphi_{i} \vert a_{m}(\varphi_{j}))$. 
$a_{n}(\varphi_{i} \vert a_{m}(\varphi_{j}))$ is defined as a mean $a_{n}(\varphi_{i})$ value for given events with $a_{m}(\varphi_{j})$. 
In other words, once events with a certain $a_{m}$ value in $\varphi_{j}$ bin are selected, the mean $a_{n}$ in the $\varphi_{i}$ bin is estimated for the value of $a_{n}(\varphi_{i} \vert a_{m}(\varphi_{j}))$. 
Various combinations of $n$, $m$, ($n, m=$1,~2,~3, ...) and $i$, $j$ ($i,j=$in, out) can be investigated to reveal features of azimuthal correlations of the longitudinal structures, but only $n=m$ cases up to $n=3$ are discussed in these proceedings. 
For $(\varphi_{i},\varphi_{j})$ combinations, four groups of conditional $a_{n}$ are studied:
\begin{itemize}
\item $a_{n}(\varphi_{\text{opp.in}} \vert a_{n}(\varphi_{\text{in}}))$ 
\item $a_{n}(\varphi_{\text{opp.out}} \vert a_{n}(\varphi_{\text{out}}))$ 
\item $a_{n}(\varphi_{\text{out}} \vert a_{n}(\varphi_{\text{in}}))$ 
\item $a_{n}(\varphi_{\text{in}} \vert a_{n}(\varphi_{\text{out}}))$ 
\end{itemize}
where ``opp.in'' and ``opp.out'' correspond to the opposite in-plane and opposite out-of-plane $\varphi$ bins, respectively. 

\section{Experimental setup}
For the current analysis, ALICE data from the Pb--Pb run at the LHC in 2011 are used.
A full description of the ALICE detector and its performance during this period can be found in~\cite{ALICE:2014p}, while a few details with direct relation to this analysis will be explained in this section.

The online event selection for the analysis combined the minimum-bias (MB) trigger and two centrality triggers, namely, central and semi-central triggers. 
Requiring signals in two scintillator arrays (V0) and Zero Degree Calorimeters (ZDC) in the MB trigger suppressed events from only electromagnetic interactions between the lead ions. 
Two centrality triggers required additional V0 signals, and selected large numbers of events particularly in central collisions. 
Events were then classified into centrality classes based on the signal in V0 detectors and 0--5\%, 5--10\%, 10--20\%, 20--30\%, 30--40\%, 40--50\% centrality classes were used for the analysis. 

As explained in Sec.~\ref{Sec:Def}, the second-order event plane, $\Psi_{2}$, should first be determined for conditional $a_{n}$ measurements. 
$\Psi_{2}$ is determined by 
\begin{eqnarray}
Q_{2,x} =  \sum\limits_{i=1}^n w_{i}\,\text{cos}(2\varphi_{i}) \; \text{,} \; \;Q_{2,y} =  \sum\limits_{i=1}^n w_{i}\,\text{sin}(2\varphi_{i}) \; \text{,}
\label{eq:Qvector}
\end{eqnarray}
and
\begin{eqnarray}
\Psi_{2}=  \frac{1}{n}\text{tan}^{-1}\left(\frac{Q_{2,y}}{Q_{2,x}} \right) \; \text{,}
\label{eq:psi2}
\end{eqnarray}
using signals from the V0 detectors, which cover $2.8 < \eta < 5.1$ and $-3.7 < \eta < -1.7$.
In Eq.~\ref{eq:Qvector}, index $i$ runs over sectors of V0 detectors, and $w_{i}$ is the measured signal in sector $i$. 
Due to the non-uniform gain of V0 detectors, the calibration for the $\Psi_{2}$ estimation was performed using the procedures described in~\cite{Poskanzer:1998yz,Selyuzhenkov:2007zi,MASSACRIER2013235}, such as a gain equalization of the V0 detector signal, a flattening of the $\Psi_{2}$ distribution through $Q$-vector re-centering, twisting, and rescaling.
Estimating $\Psi_{2}$ with the V0 detectors ensures a sufficient $\eta$ gap between ranges of $\Psi_{2}$ and conditional $a_{n}$ measurements, as mid-rapidity tracks are used for the conditional $a_{n}$.
Such a separation allows to avoid auto-correlation biases between $\Psi_{2}$ and tracks in the mid-rapidity.

The mid-rapidity charged tracks are reconstructed using clusters of the inner tracking system (ITS) and the time projection chamber (TPC) in order to achieve uniform tracking efficiency~\cite{ALME2010316}. 
The default track selection required at least 70 space points in the TPC, the ratio of the number of recorded points to the findable points larger than 80\%, a distance of closest approach (DCA) to the reconstructed vertex smaller than 2.4\,cm in the transverse direction and 3.2\,cm in the longitudinal direction, and the maximum fraction of shared clusters in the TPC lower than 40\%. 
In the process of matching TPC tracks with ITS clusters, two approaches were followed: tracks reconstructed with at least one hit in the SPD, and tracks without associated SPD hit but the reconstructed vertex position in the fit. 
The $p_{\text{T}}$ and $\eta$ ranges of tracks are $0.2 < p_{\text{T}} < 5.0$\,GeV/$c$ and $-0.8 < \eta < 0.8$, respectively.

\section{Results}
In the measuremnts of the conditional $a_{n}$, two correction procedures are applied to take into account detector inefficiencies. 
First, an event-plane resolution correction accounts for the difference between the true second-order event plane and the measured second-order event plane. 
The corresponding correction procedure in the flow measurements in the transverse direction is well-established~\cite{Poskanzer:1998yz}. 
Another correction accounts for the multiplicity effects due to the limited tracking efficiency in the ITS and TPC. 
For a given true $a_{n}$ value, the distribution of measured $a_{n}$ depends on the charged-particle multiplicity used in the $a_{n}$ measurement.
This dependence is investigated with the Monte Carlo method, which simulates the response in the measured $a_{n}$ from an input $a_{n}$ with various multiplicity distributions. 
Deviations attributed to the difference in the measured multiplicity from the true multiplicity in observables are corrected, accordingly. 

For each conditional $a_{n}$ result, systematic uncertainties are assigned by varying the measurement conditions and correction details. 
These include centrality estimation methods, detector ranges for the event-plane determination, selected ranges of reconstructed primary vertex coordinate along the beam direction, track selection, functional forms in the event-plane resolution correction, and variationss in the correction for the multiplicity effects. 
The total systematic uncertainty is evaluated by combining deviations to the values from the nominal measurement in each variation quadratically. 

\begin{figure}
	\centering
	\includegraphics[width=0.4\textwidth]{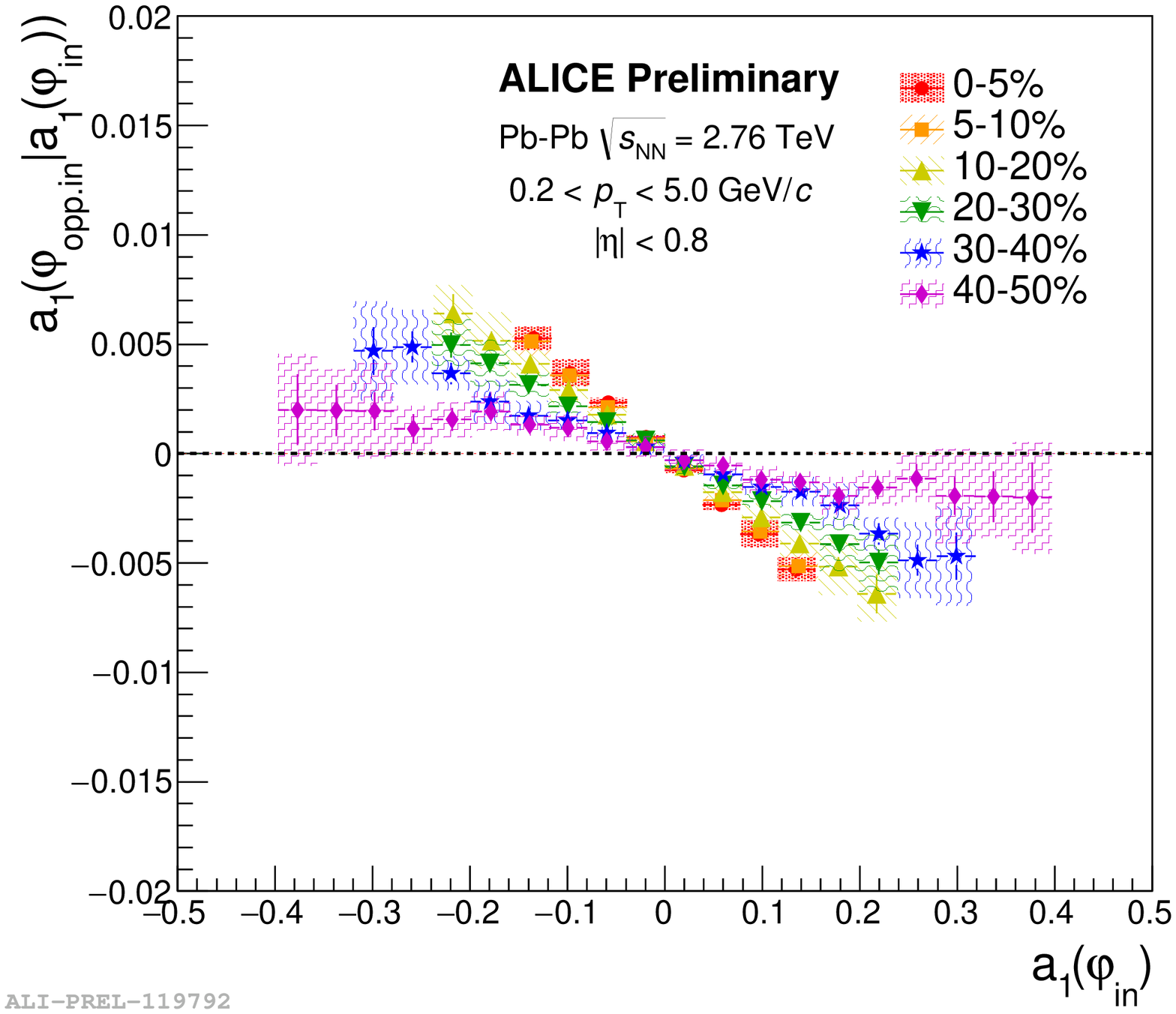}
	\includegraphics[width=0.4\textwidth]{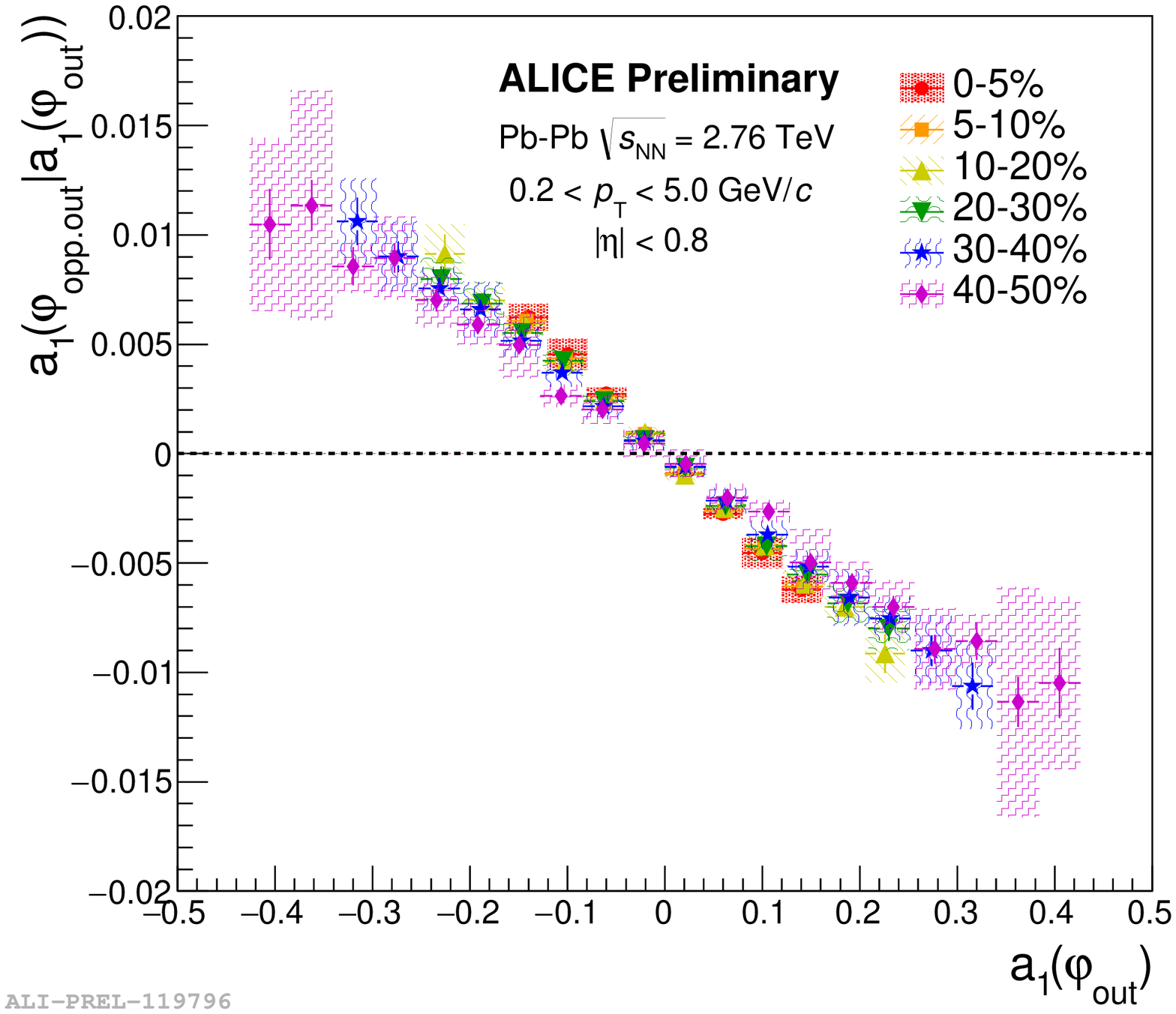}
	\includegraphics[width=0.4\textwidth]{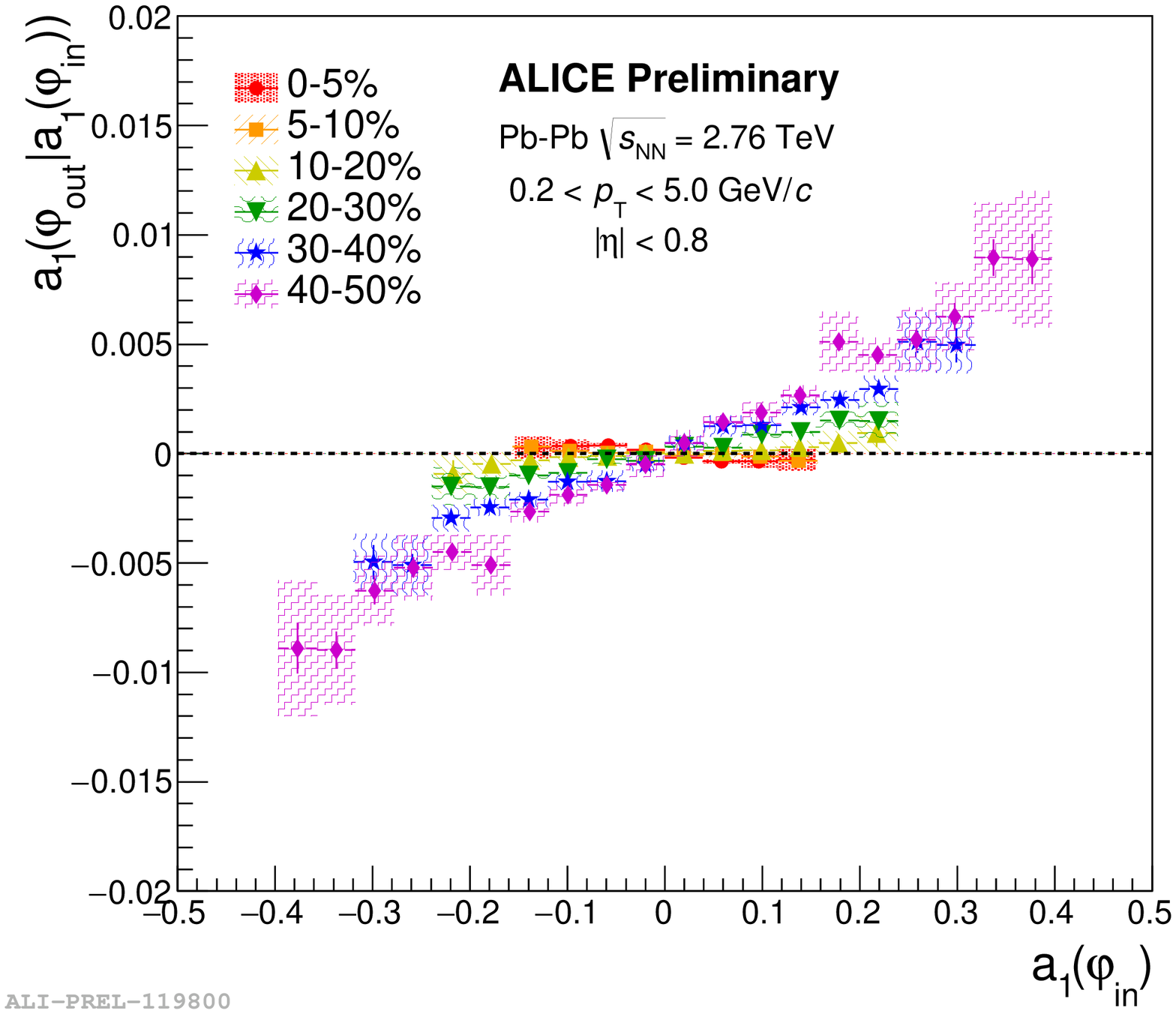}
	\includegraphics[width=0.4\textwidth]{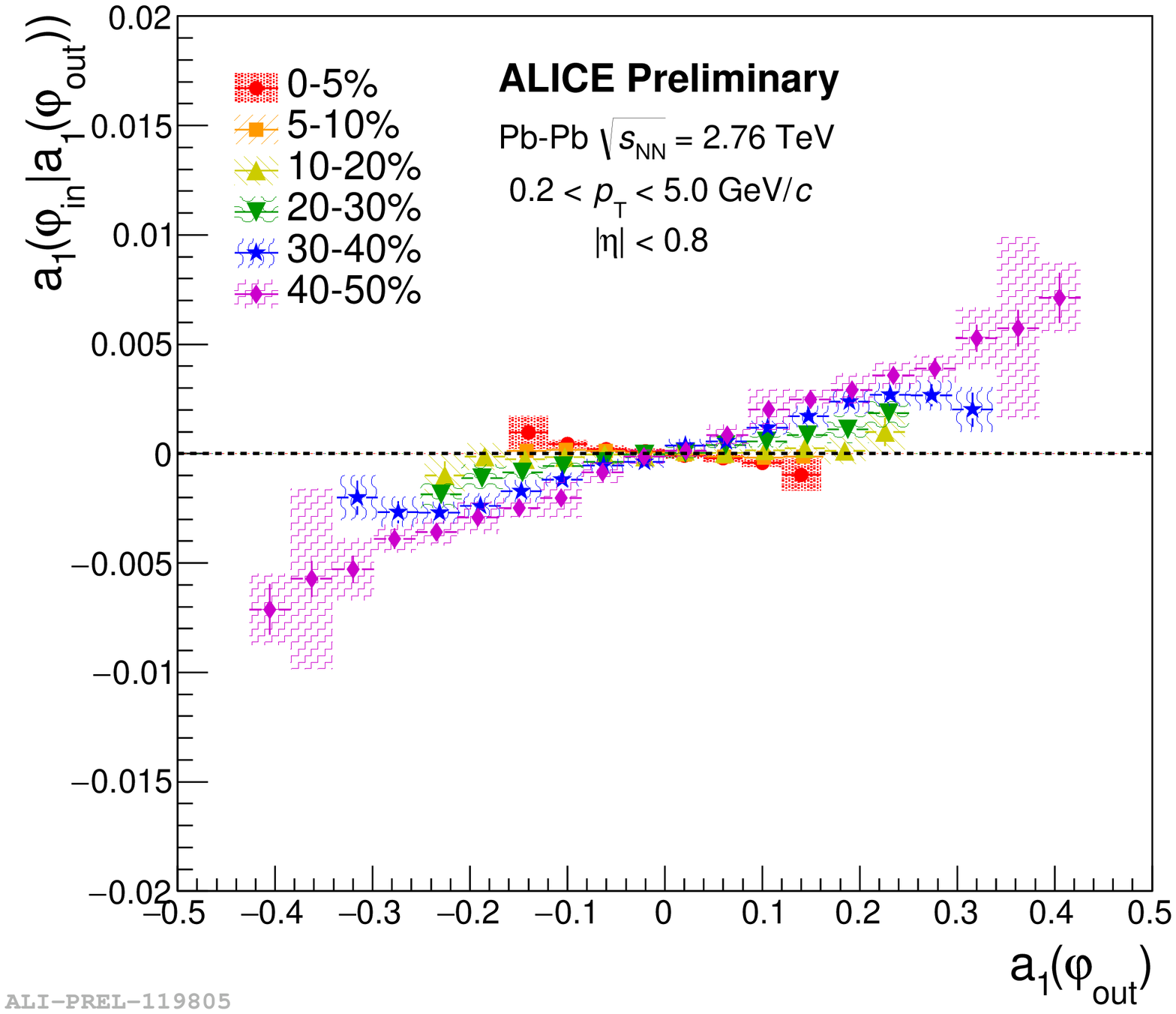}
	\caption{\label{fig:Sys_a1} The $a_{1}(\varphi_{\text{opp.in}} \vert a_{1}(\varphi_{\text{in}}))$ (top left), $a_{1}(\varphi_{\text{opp.out}} \vert a_{1}(\varphi_{\text{out}}))$ (top right), $a_{1}(\varphi_{\text{out}} \vert a_{1}(\varphi_{\text{in}}))$ (bottom left), and $a_{1}(\varphi_{\text{in}} \vert a_{1}(\varphi_{\text{out}}))$ (bottom right) measured with charged particles with $0.2 < p_{\text{T}} < 5.0$\,GeV/$c$ and $\vert \eta \vert < 0.8$ in different centrality classes.}
\end{figure}

The conditional $a_{1}$ of four azimuthal combinations for different centrality classes are shown in Fig.~\ref{fig:Sys_a1}. 
Statistical uncertainties are marked with error bars and systematic uncertainties are marked with shaded areas in every figure.
\begin{figure}
	\centering
	\includegraphics[width=0.29\textwidth]{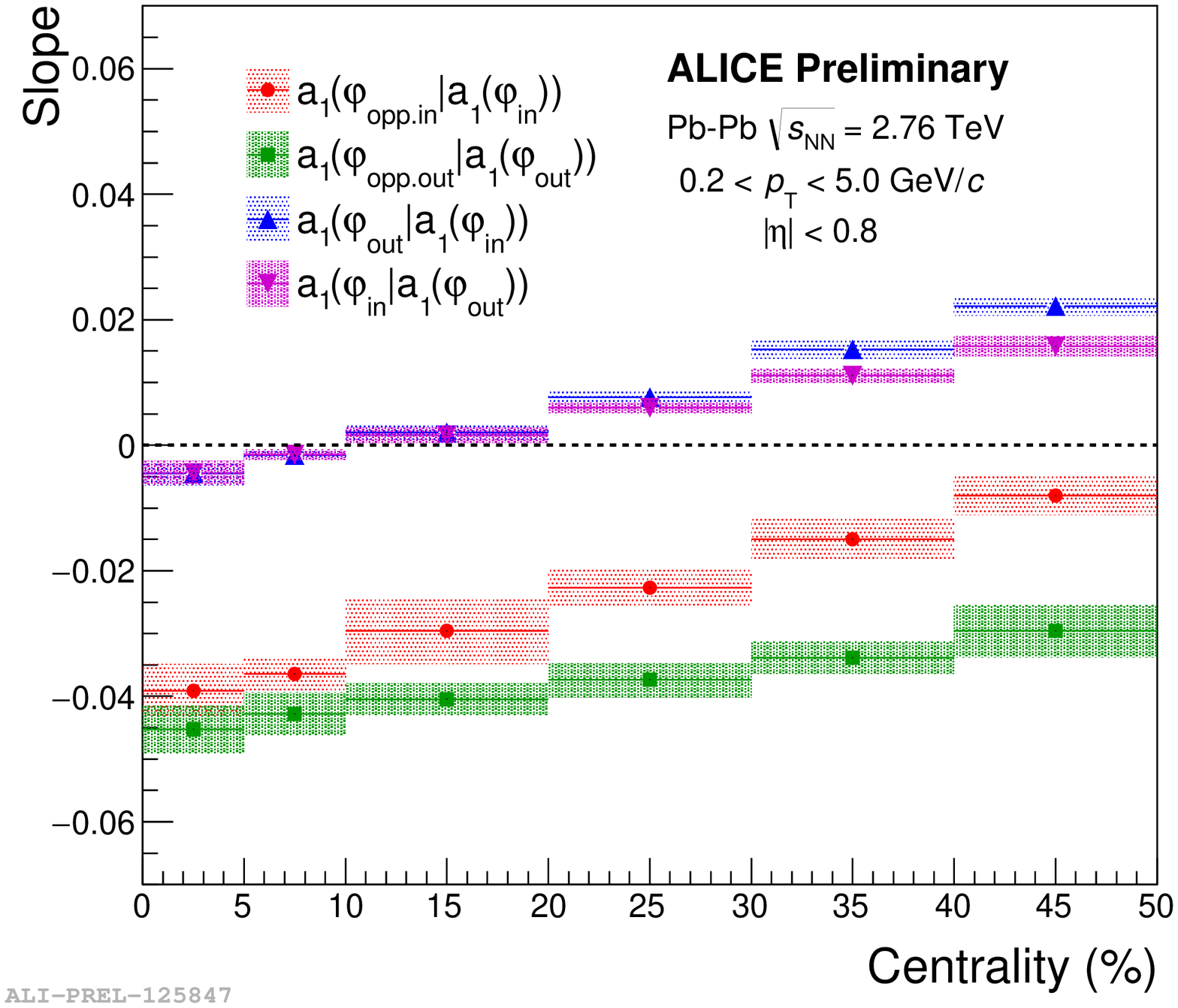}\hspace{2pc}
	\includegraphics[width=0.29\textwidth]{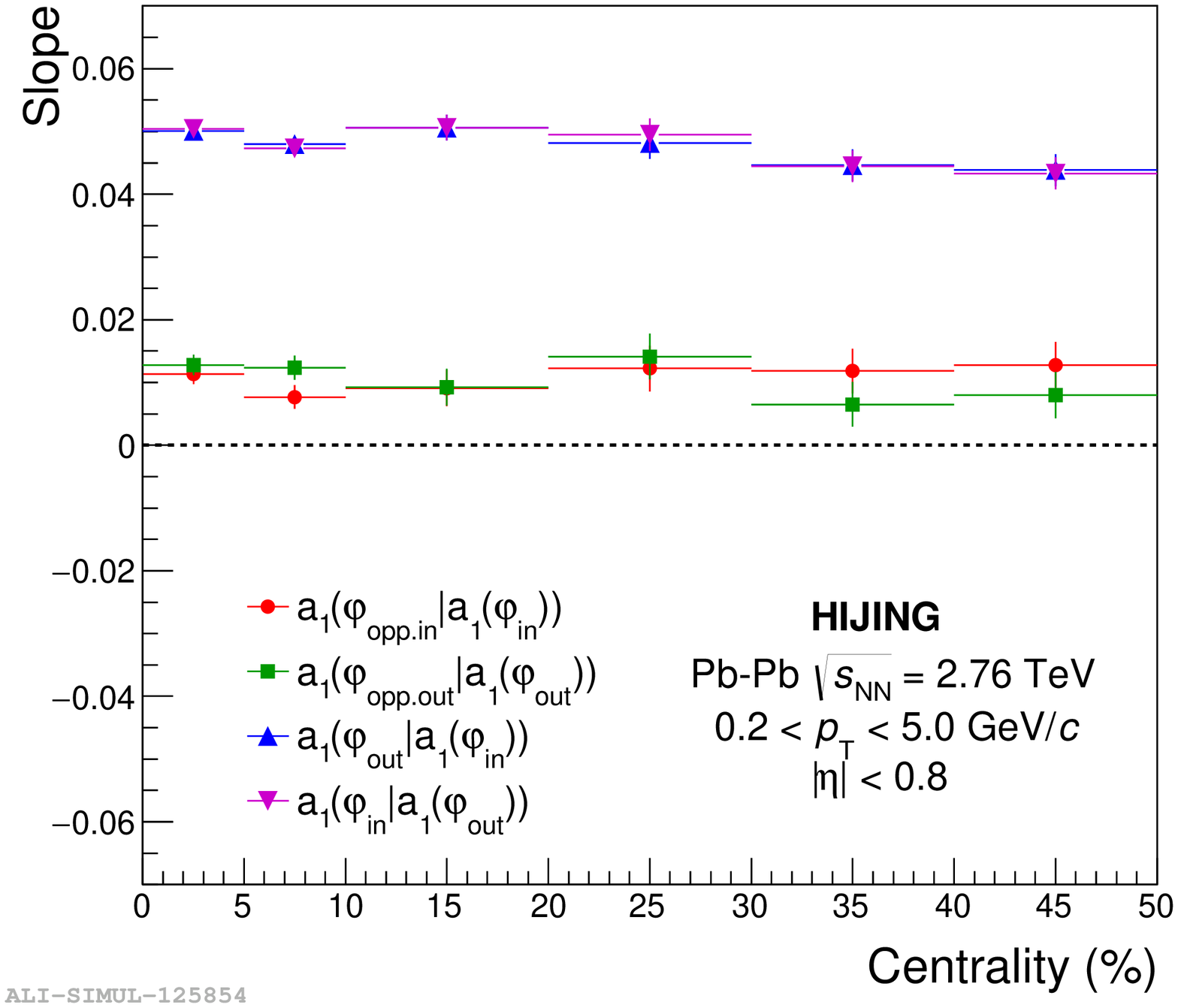}\hspace{2pc}	
	\includegraphics[width=0.29\textwidth]{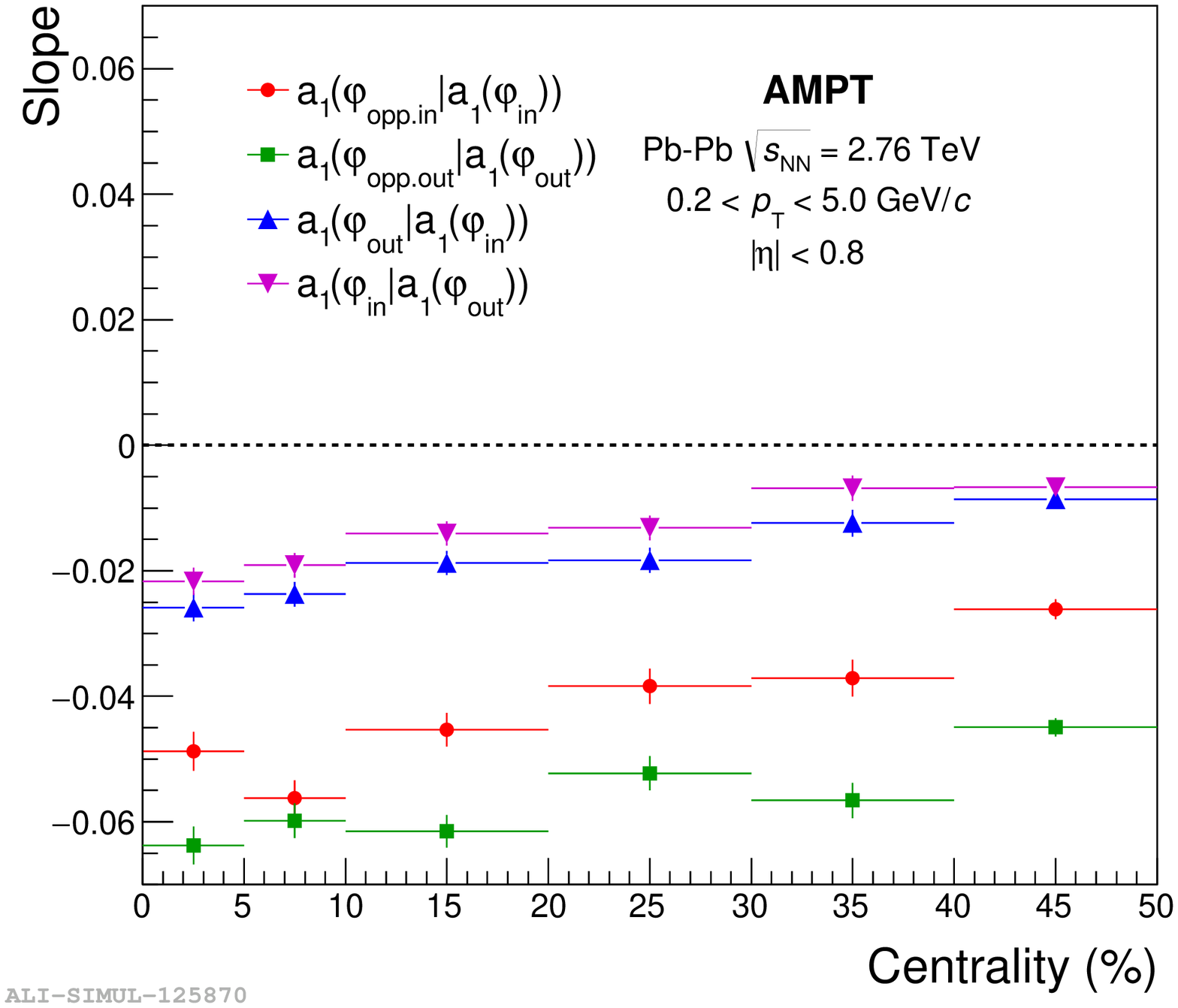}\hspace{2pc}
	\caption{\label{fig:Sys_anSlope} Slopes of conditional $a_{1}$ as a function of centrality from ALICE measurements at the LHC (left), HIJING (center), and AMPT (right) simulations.}
\end{figure}
These conditional $a_{1}$ results are fitted with a linear function, and the extracted slopes are plotted in Fig.~\ref{fig:Sys_anSlope} along with the quantities obtained with HIJING~\cite{Wang:1991hta} and AMPT~\cite{Lin:2004en} simulations. 

The $a_{1}$ is a parameter quantifying the forward-backward asymmetry in particle production, and the plots in Fig.~\ref{fig:Sys_a1} and Fig.~\ref{fig:Sys_anSlope} show a distinctive level of contributions to the longitudinal structure from collective and non-collective correlations depending on the centrality.  
In general, physics mechanisms studied via geometrical correlations can be classified into two groups, collective and non-collective correlations.
Particles originating from non-collective correlations, such as jets and resonances, are localized in a small range of ($\varphi$, $\eta$) space or opposite $\varphi$ regions, while particles from collective correlations, i.e. flow, are distributed over a wide range in ($\varphi, \eta$) space.  
Due to the limited range of non-collective correlations in $(\varphi, \eta)$ space, non-collective correlations can contribute to either one or two adjacent $\varphi$ bins in the current analysis for most cases, as the size of each $\varphi$ bin is $\pi/2$. 
For example, fragmented particles from a jet can be included in either a single $\varphi$ bin or two adjacent $\varphi$ bins. 
In the case that the jet axis is located in the positive $\eta$ region, the corresponding fragmented particles can contribute to the positive $a_{1}$ in a given $\varphi$ bin or two adjacent $\varphi$ bins simultaneously. 
More positive values of slopes in $a_{1}(\varphi_{\text{out}} \vert a_{1}(\varphi_{\text{in}}))$ and $a_{1}(\varphi_{\text{in}} \vert a_{1}(\varphi_{\text{out}}))$ than those in $a_{1}(\varphi_{\text{opp.in}} \vert a_{1}(\varphi_{\text{in}}))$ and $a_{1}(\varphi_{\text{opp.out}} \vert a_{1}(\varphi_{\text{out}}))$ are mostly attributed to the involvement of the non-collective correlations in the observables, and indicate larger contributions in more peripheral classes. 
Comparison with HIJING results on Fig.~\ref{fig:Sys_anSlope}, where no collective correlations are expected, indicates larger contributions from collective correlations in more central classes. 

Collective correlations contribute to the conditional $a_{1}$ for all $(\varphi_{i}, \varphi_{j})$ combinations. 
Within the framework of collective correlations, various physics scenarios can result in the negative slopes observed in $a_{1}(\varphi_{\text{opp.in}} \vert a_{1}(\varphi_{\text{in}}))$ and $a_{1}(\varphi_{\text{opp.out}} \vert a_{1}(\varphi_{\text{out}}))$. 
The transverse distribution of the asymmetric number of participating nucleons from two incoming nuclei fluctuates event-to-event, and can be considered as a seed for the asymmetric expansion of the medium in the longitudinal direction. 
Then the negative slope in conditional $a_{1}$ with two opposite $\varphi$ bins may originate solely from the initial bias in the transverse distribution of the asymmetric number of participating nucleons, although this initial bias may be diluted through the expansion. 
In the same sense, the difference between $a_{1}(\varphi_{\text{opp.in}} \vert a_{1}(\varphi_{\text{in}}))$ and $a_{1}(\varphi_{\text{opp.out}} \vert a_{1}(\varphi_{\text{out}}))$ may attribute to the different expansion rate in in-plane and out-of-plane directions, as the difference is the smallest in the most central class. 
Also, the momentum conservation of particles created from the expansion of the medium may impact these results. 
Slopes of conditional $a_{1}$ in AMPT are observed to have similar centrality dependence (Fig.~\ref{fig:Sys_anSlope}), but generally shifted in the negative direction. 
This can be interpreted as a different level of contributions from collective and non-collective correlations to the data.

\begin{figure}
	\centering
	\includegraphics[width=0.4\textwidth]{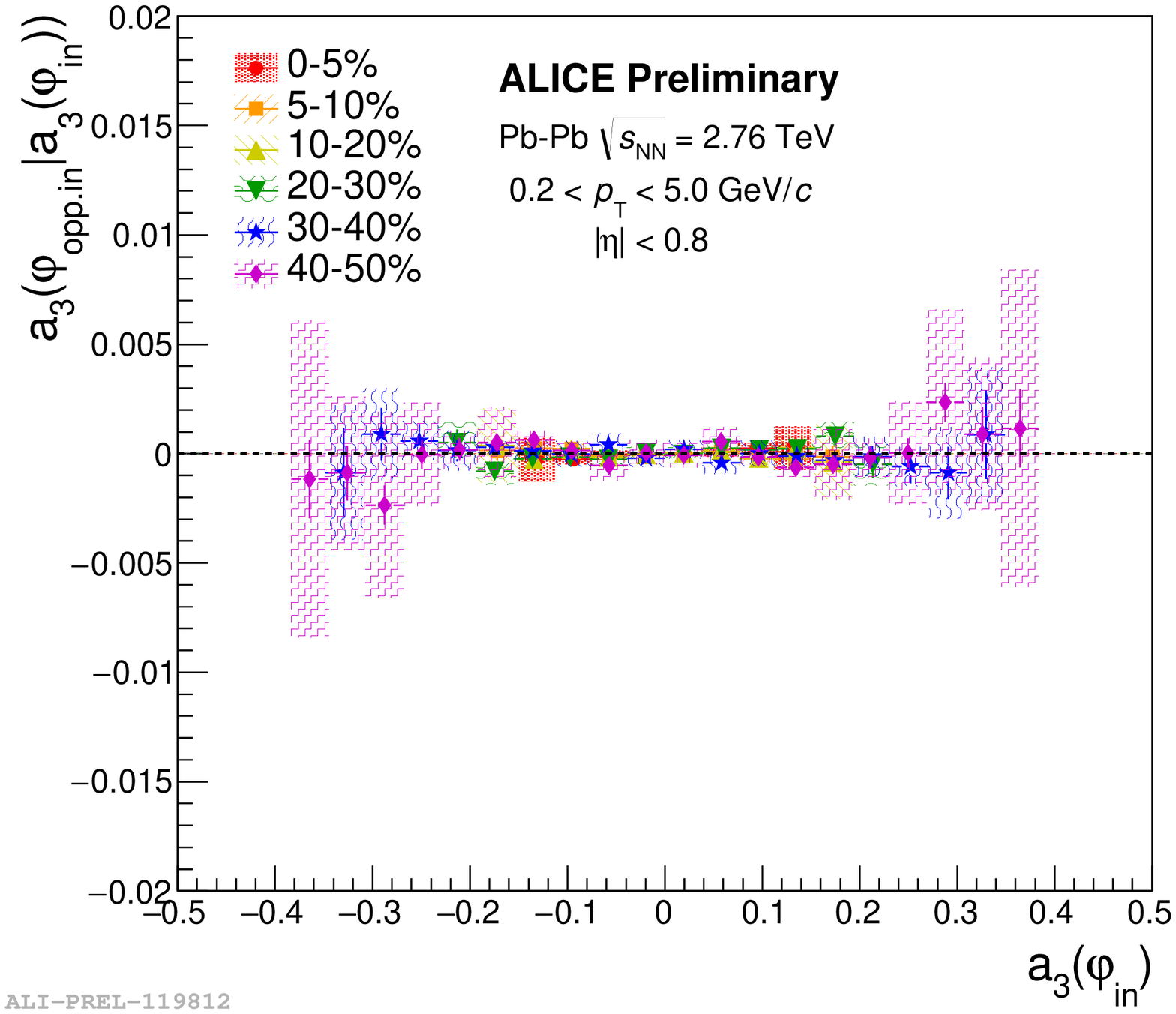}
	\includegraphics[width=0.4\textwidth]{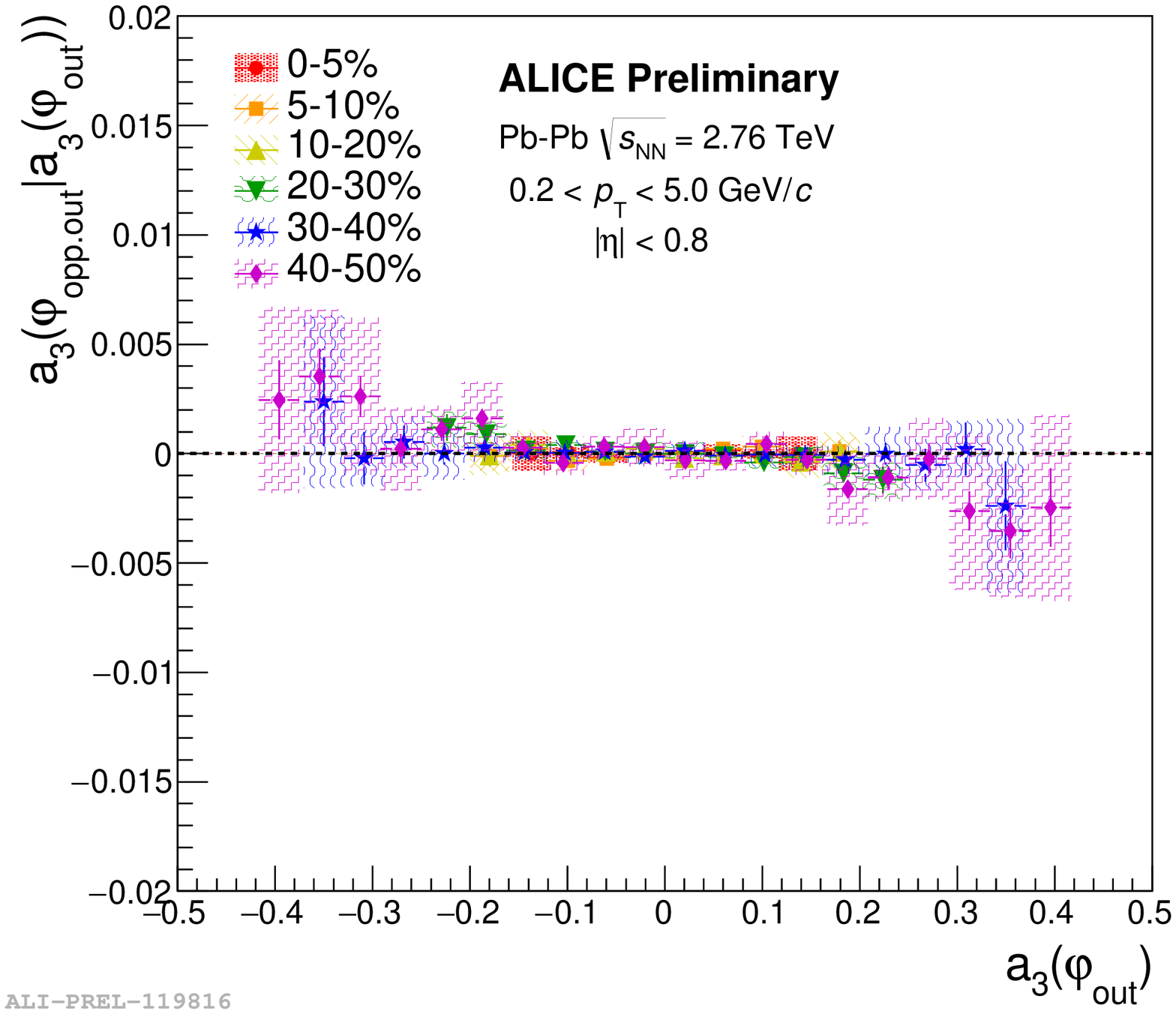}
	\includegraphics[width=0.4\textwidth]{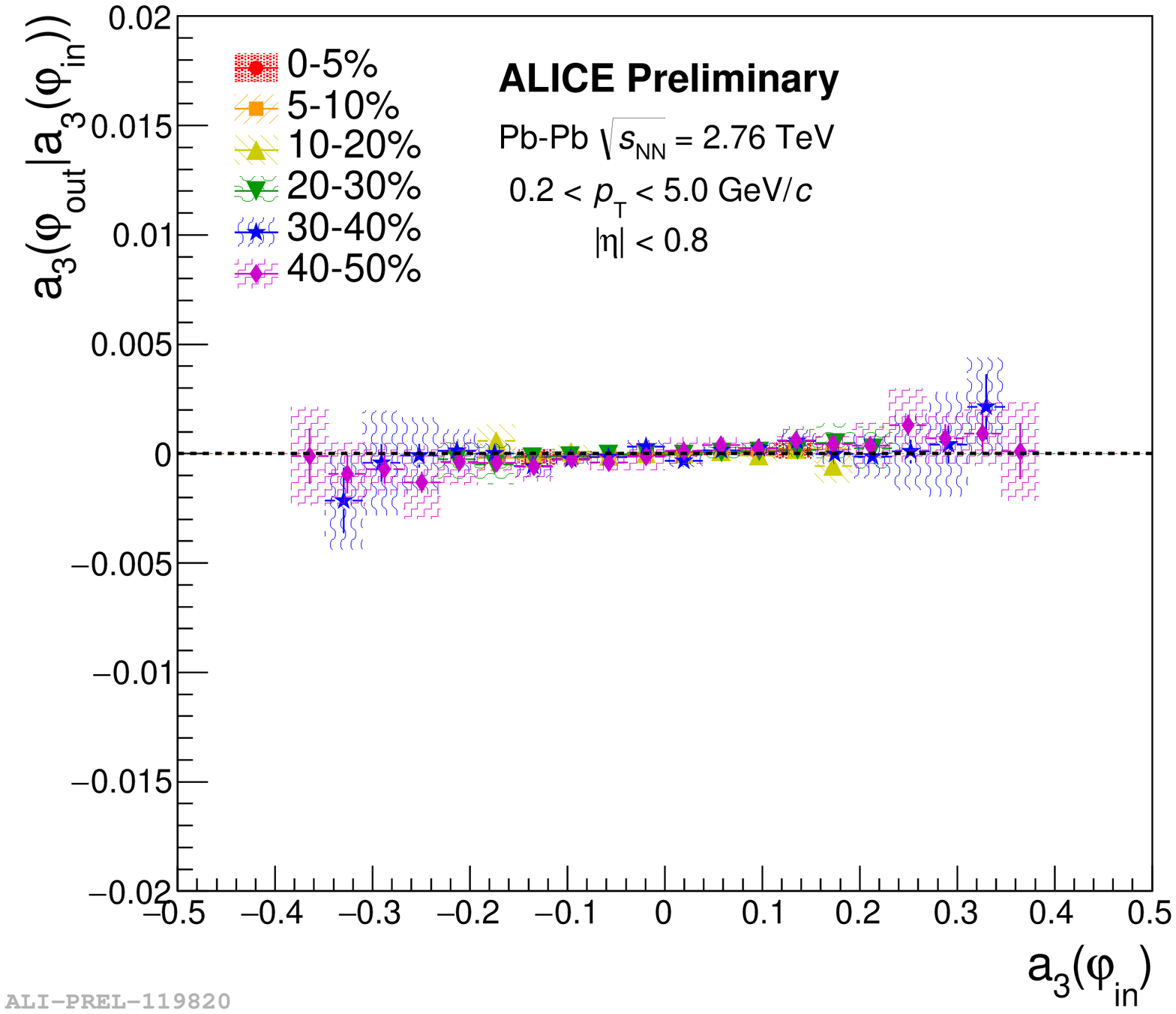}
	\includegraphics[width=0.4\textwidth]{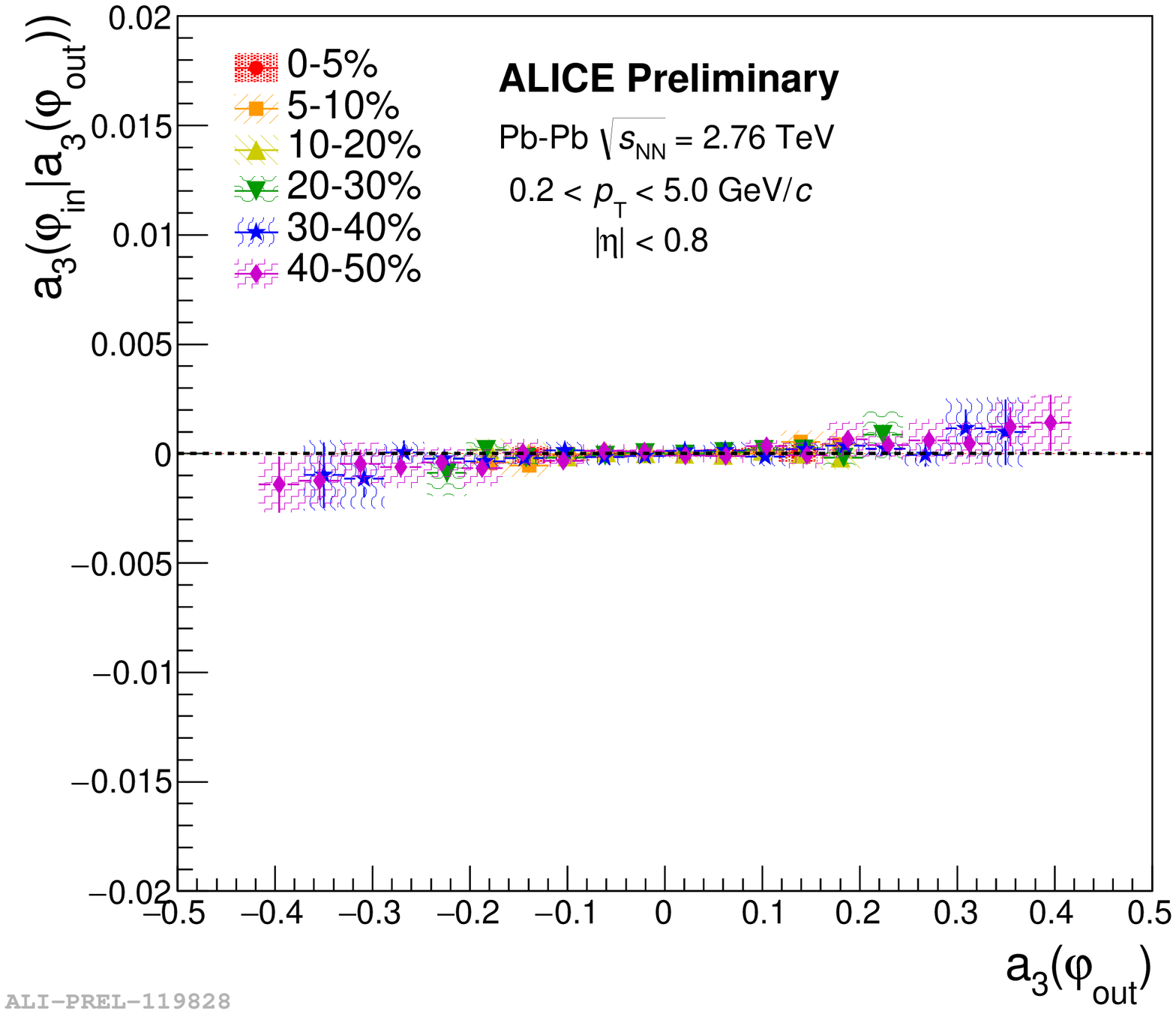}
	\caption{\label{fig:Sys_a3} The $a_{3}(\varphi_{\text{opp.in}} \vert a_{3}(\varphi_{\text{in}}))$ (top left), $a_{3}(\varphi_{\text{opp.out}} \vert a_{3}(\varphi_{\text{out}}))$ (top right), $a_{3}(\varphi_{\text{out}} \vert a_{3}(\varphi_{\text{in}}))$ (bottom left), and $a_{3}(\varphi_{\text{in}} \vert a_{3}(\varphi_{\text{out}}))$ (bottom right) measured with charged particles with $0.2 < p_{\text{T}} < 5.0$\,GeV/$c$ and $\vert \eta \vert < 0.8$ in different centrality classes.}
\end{figure}

In~\cite{Bzdak:2013aa,ATLAS:2017aa}, $a_{3}$ and higher order coefficients are argued to mostly attribute to the non-collective or short-range correlations. 
The conditional $a_{3}$ results for different centrality classes are shown in Fig.~\ref{fig:Sys_a3}. 
The conditional $a_{3}$ values remain near 0 in all $(\varphi_{i}, \varphi_{j})$ combinations, which indicates independence of $a_{3}$ in different $\varphi$ regions. 
This is a genuine feature of non-collective correlations, as non-collective correlations can generally contribute to the local region in $(\varphi, \eta)$ space. 
Similar features were found in HIJING and AMPT simulations although they are now shown in these proceedings. 

\begin{figure}
	\centering
	\includegraphics[width=0.4\textwidth]{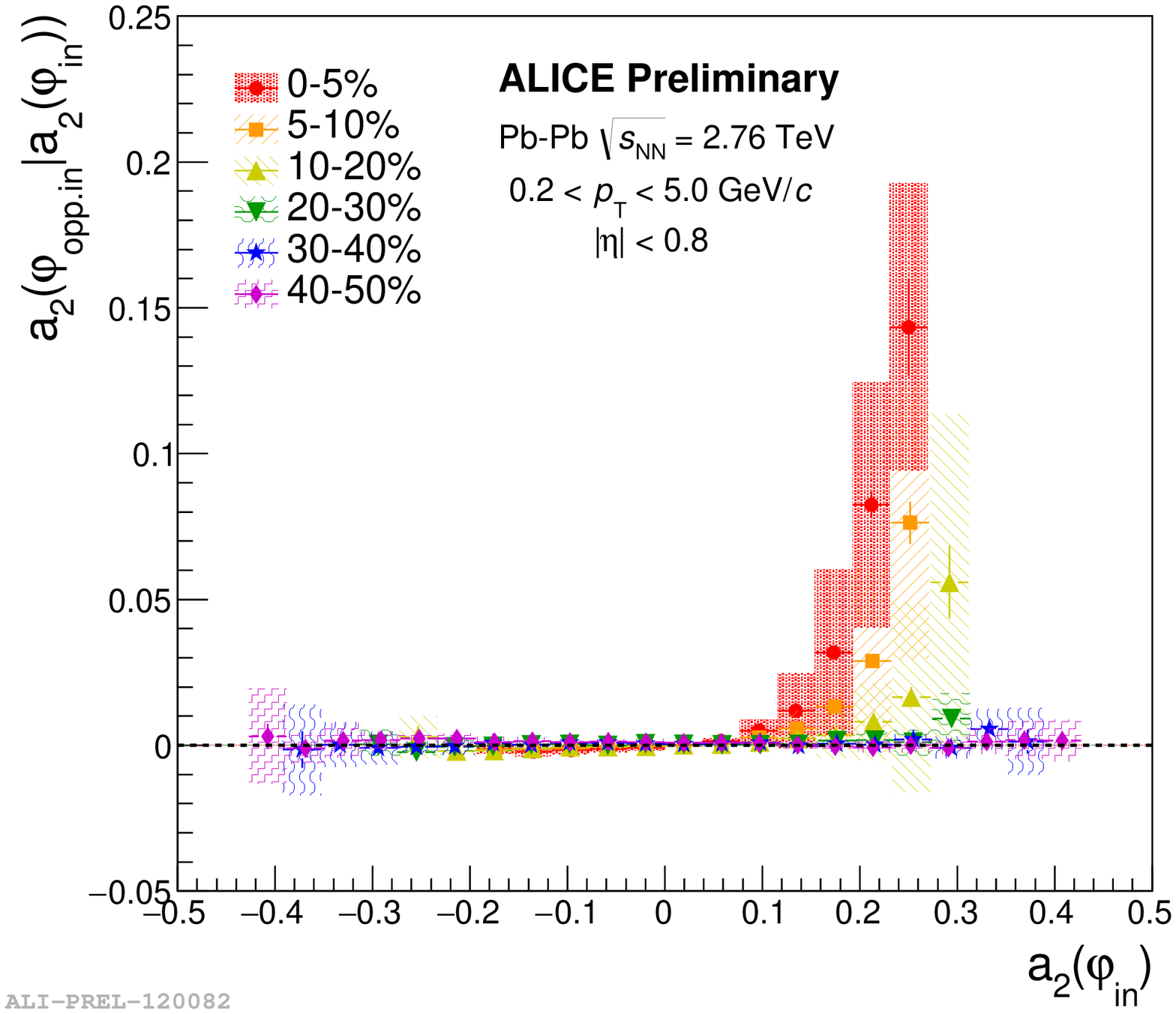}
	\includegraphics[width=0.4\textwidth]{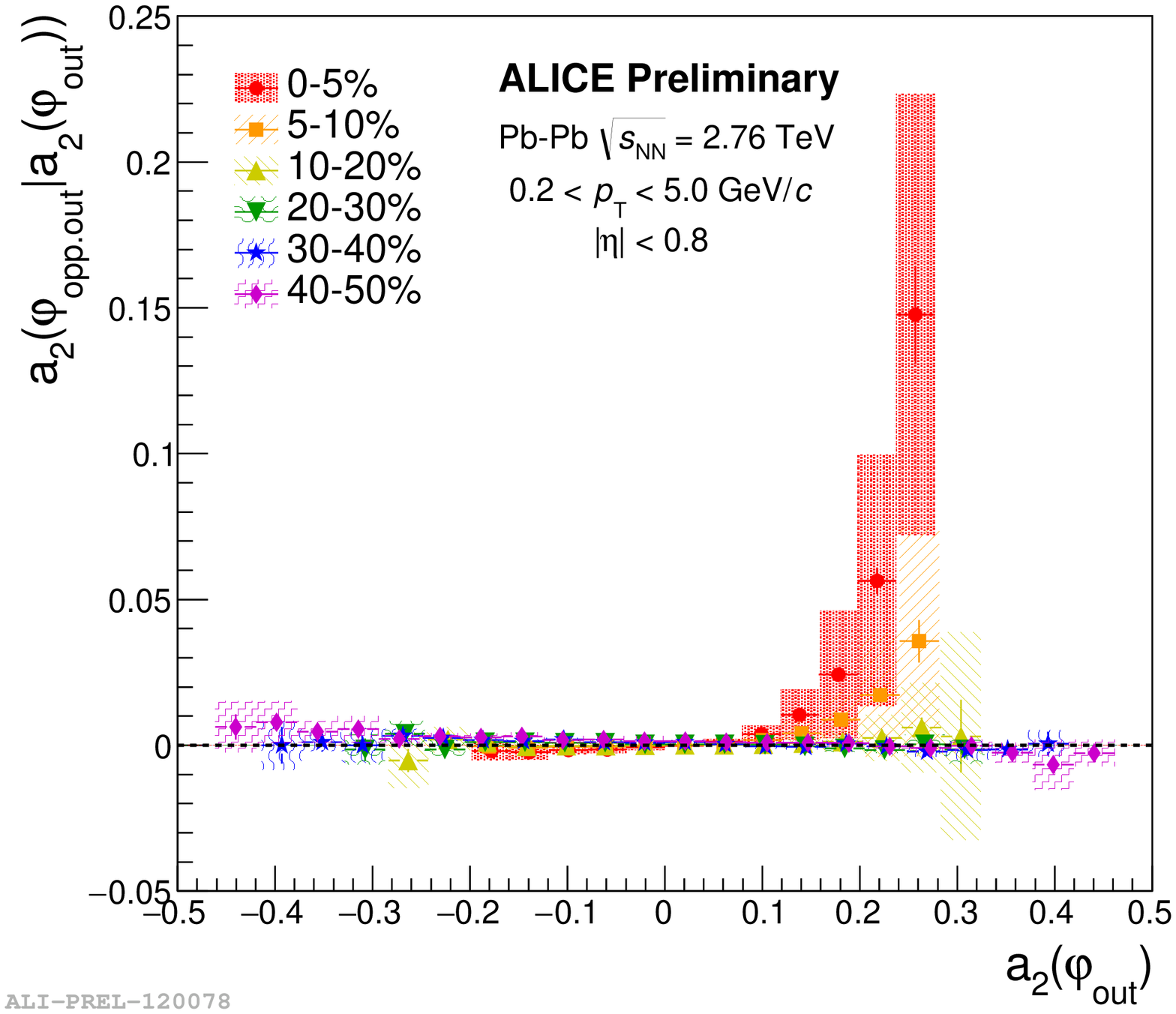}
	\includegraphics[width=0.4\textwidth]{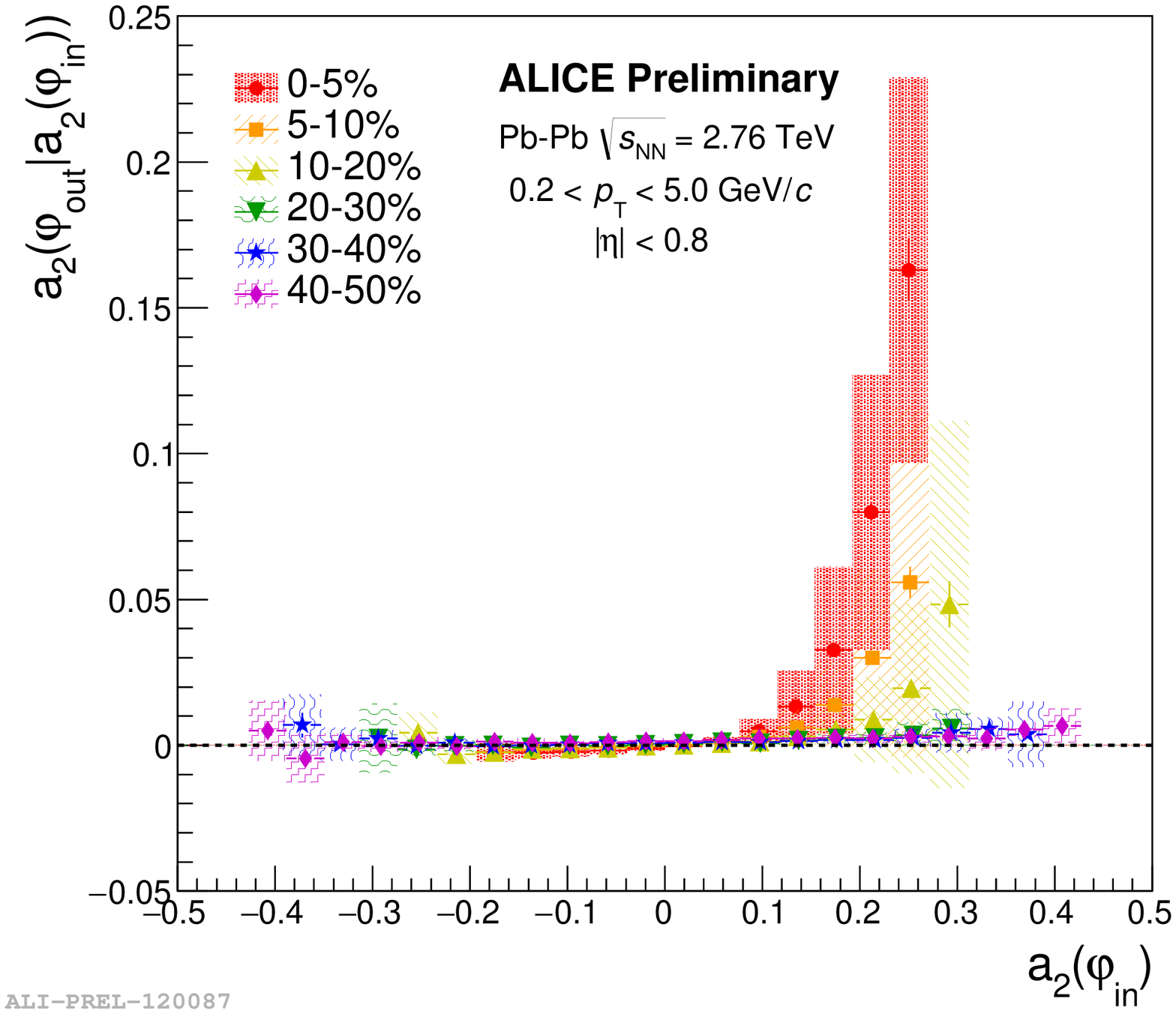}
	\includegraphics[width=0.4\textwidth]{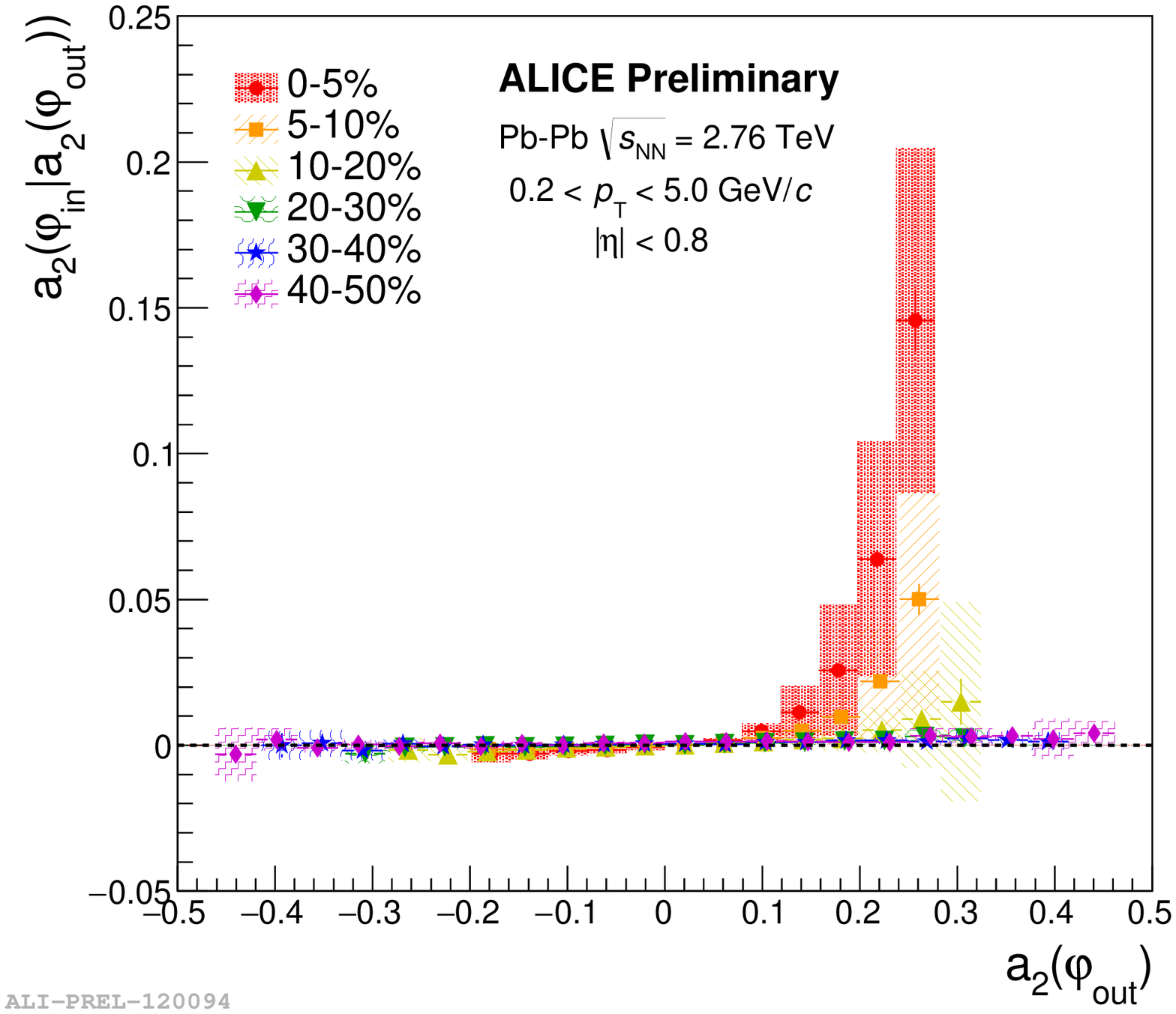}
	\caption{\label{fig:Sys_a2} The $a_{2}(\varphi_{\text{opp.in}} \vert a_{2}(\varphi_{\text{in}}))$ (top left), $a_{2}(\varphi_{\text{opp.out}} \vert a_{2}(\varphi_{\text{out}}))$ (top right), $a_{2}(\varphi_{\text{out}} \vert a_{2}(\varphi_{\text{in}}))$ (bottom left), and $a_{2}(\varphi_{\text{in}} \vert a_{2}(\varphi_{\text{out}}))$ (bottom right) measured with charged particles with $0.2 < p_{\text{T}} < 5.0$\,GeV/$c$ and $\vert \eta \vert < 0.8$ in different centrality classes.}
\end{figure}

Lastly, the conditional $a_{2}$ measurements for different centrality classes are shown in Fig.~\ref{fig:Sys_a2}. 
The $a_{2}$ is a parameter quantifying the mid-peripheral asymmetry, and related to the amount of nuclear stopping power or shift of the effective center-of-mass of the collisions~\cite{ATLAS:2017aa}. 
A positive $a_{2}$ represents more particles in peripheral regions than the central region, and a negative $a_{2}$ represents more particles in the central region than peripheral regions. 
Figure~\ref{fig:Sys_a2} shows a very distinctive feature of conditional $a_{2}$ relative to conditional $a_{1}$ or conditional $a_{3}$. 
For all $(\varphi_{i}, \varphi_{j})$ combinations, the value of $a_{2}(\varphi_{i} \vert a_{2}(\varphi_{j}))$ significantly increases for positive $a_{2}(\varphi_{j})$ in central classes, while it remains near 0 for other $a_{2}(\varphi_{j})$.
This indicates that when a positive value of $a_{2}$ is observed in a given $\varphi$ bin in central collisions, values of $a_{2}$ in other $\varphi$ bins tend to be positive. 
Such collective features disappear in more peripheral classes, and are not observed in HIJING and AMPT simulations. 

\section{Summary}
The ALICE Collaboration reports the first results of azimuthal correlations of longitudinal structure of charged-particle multiplicity in Pb-Pb collisions with newly devised observables, conditional $a_{n}$. 
While previous measurements in the longitudinal direction are limited to exclusively dealing with the pseudorapidity of produced particles, these new measurements extend the scope of longitudinal correlations into a 3-dimensional space. 
The conditional $a_{1}$ results show a different level of contributions from collective and non-collective correlations into a forward-backward asymmetry depending on the centrality. 
The conditional $a_{3}$ results confirm the origin of $a_{3}$, which is expected to be due to non-collective correlations. 
A unique collective feature in central classes is observed in conditional $a_{2}$, which indicates that a positive $a_{2}$ is preserved over the azimuthal angle in more central classes.  
Both HIJING and AMPT are not able to reproduce such features. 
These results represent new constraints for the 3-dimensional heavy-ion collision models, and require further theoretical investigation.

\section*{Acknowledgements}
This work is supported by the U.S. Department of Energy under grant number DE-SC0004168.

\section*{References}
\bibliography{iopart-num}

\end{document}